\documentclass[fleqn,12pt]{wlscirep}

\makeatletter
\newcommand*\bigcdot{\mathpalette\bigcdot@{.5}}
\newcommand*\bigcdot@[2]{\mathbin{\vcenter{\hbox{\scalebox{#2}{$\m@th#1\bullet$}}}}}
\makeatother

\usepackage{enumitem}
\usepackage{amsmath,amssymb}

\title{Optimal control of the transport of Bose-Einstein condensates with atom chips}

\author[1,2]{S. Amri}
\author[1,2]{R. Corgier}
\author[3]{D. Sugny}
\author[2]{E. M. Rasel}
\author[2]{N. Gaaloul}
\author[1]{E. Charron}

\affil[1]{ISMO, CNRS, Univ. Paris-Sud, Universit\'e Paris-Saclay, F-91405, Orsay cedex, France}

\affil[2]{Institut f\"ur Quantenoptik, Leibniz Universit\"at Hannover, Welfengarten 1, D-30167 Hannover, Germany}

\affil[3]{LICB, CNRS, Université Bourgogne Franche Comté, 9 Av. A. Savary, BP 47870, F-21078 Dijon cedex, France}

\begin{abstract}
Using Optimal Control Theory (OCT), we design fast ramps for the controlled transport of Bose-Einstein condensates with atom chips' magnetic traps. These ramps are engineered in the context of precision atom interferometry experiments and support transport over large distances, typically of the order of 1\,mm, \emph{i.e.} about 1,000 times the size of the atomic clouds, yet with durations not exceeding 200\,ms. We show that with such transport durations of the order of the trap period, one can recover the ground state of the final trap at the end of the transport. The performance of the OCT procedure is compared to that of a Shortcut-To-Adiabaticity (STA) protocol and the respective advantages / disadvantages of the OCT treatment over the STA one are discussed.
\end{abstract}

\begin{document}

\flushbottom
\maketitle

\thispagestyle{empty}

\section*{Introduction}

The measurement's outcome of a phase-sensitive sensor probing forces exerted on neutral atoms by inertial, material or electromagnetic sources depends dramatically on the initial conditions, \emph{i.e.} on the position, velocity and size of the input matter-wave. A lack of knowledge or scattering in these initial properties inevitably leads to systematic effects or statistical errors harming the sensor's performance. An example of the degree of control needed can be grasped if one considers making a test of the Universality of Free Fall (UFF) with two different atomic species to put bounds on a possible violation of the UFF at the femto-level in the Eötvös ratio \cite{Will2014}, level at which state-of-the-art experiments perform with material test masses \cite{MicroscopePRL2017}. Such a precise experiment requires that the initial positions, center-of-mass velocities and expansion rates are defined at a level better than $1\,\mu$m, $1\,\mu$m/s and $100\,\mu$m/s (35 pK in 3D), respectively \cite{HartwigNJP2015}.

To meet these stringent requirements, the temperatures of the atomic ensembles have to be drastically reduced (down to a sub-nK level) and their size must remain compact (not exceeding a few mm after several seconds of free expansion) clearly indicating the necessity of using Bose-Einstein Condensates (BEC). Such a direction is taken by several metrology groups worldwide \cite{DickersonPRL2013,AltinNJP2013,HardmanPRL2016,OverstreetPRL2018,AlauzeNJP2018,KarcherNJP2018,Plotkin-SwingPRL2018}, including the QUANTUS and MAIUS consortia \cite{QUANTUS-MAIUS} which reached important milestones in controlling quantum gases dynamics in microgravity conditions using atom chips \cite{MuentingaPRL2013,BeckerNature2018}.

In a recent work \cite{Corgier2018}, we considered an approach based on Shortcut-To-Adiabaticity (STA) protocols to obtain analytic solutions for the transport of the BEC in an atom chip setup with realistic anharmonic and rotating trapping potentials. This approach based on the reverse engineering technique allows for a full control of the translational degrees of freedom of the BEC. It is, however, exciting several collective modes of the quantum gas, an effect which could eventually compromise the expected metrological gain if such a source is used without any precaution as an input of an atom interferometer. It is in this context that the use of optimal control theory (OCT) can reveal an unchallenged potential of targeting a given final state in timescales shorter than the trivial adiabatic manipulation, which is of no practical use in the metrology context since it is associated with poor cycling rates.

The aim of optimal control theory is to bring a dynamical system from one state to another, while minimizing a cost functional, such as the control time or the energy of the pulse used. The modern version of OCT is born with the Pontryagin’s Maximum Principle (PMP) in the late 1950s \cite{Pontryagin1964, Lee1967}. Originally applied to problems of space dynamics, OCT is nowadays a key tool to study a large spectrum of applications both in classical \cite{Kirk2004, Bryson1975} and quantum physics \cite{Bonnard2012, Alessandro2008, Glaser2015}. In the Pontryagin formulation, solving an optimal control problem is equivalent to finding extremal trajectories which are solutions of a generalized Hamiltonian system. These trajectories satisfy the maximization condition of the PMP as well as specific boundary conditions  \cite{Kirk2004, Bryson1975, Bonnard2012}. The implementation of the PMP is far from being trivial and numerical control algorithms have been developed to approximate the optimal solution \cite{Krotov1996}. Among others, we can mention the gradient  \cite{Bryson1975, KHANEJA2005296} and the Krotov \cite{Krotov1996, 0953-4075-40-18-R01} algorithms, which are nowadays standard tools in physics.

OCT has been applied with success to quantum systems since the 1980s in domains extending from molecular physics and nuclear magnetic resonance to quantum information science (see references [\citenum{Glaser2015}] and [\citenum{1367-2630-12-7-075008}] for recent reviews, and references therein). The application of OCT to BEC dynamics has also been explored in different contexts. Using the Gross-Pitaevskii equation, the optimal coherent manipulation of an atomic BEC has been investigated in a series of studies (see references [\citenum{vFrank2016, PhysRevA.90.033628, PhysRevA.93.053612, 1367-2630-17-11-113027, PhysRevA.92.053632, PhysRevA.75.023602,AndersonPRL2018}], to cite a few, and references therein). The transport of cold atoms has also been optimized for simple models in combination with invariant-based inverse methods \cite{Corgier2018, PhysRevA.84.043415, PhysRevA.89.023627, PhysRevA.90.063425, PhysRevLett.104.063002}. It should be mentioned here that OCT and STA are usually compatible in the sense that an OCT methodology can be built on top of a basic STA frame of solutions \cite{PhysRevA.84.043415, NJP.14.013031, PRA.82.063422, Zhang_2016, PRA.85.033605, JPB.50.175501}. One can also note that recently, new methods have been tested successfully to bridge the gap between an ideal STA and a realistic experimental implementation for the optical transfer of a degenerate gas, demonstrating fast highly non-adiabatic transfer with almost no residual sloshing using corrected STA trajectories \cite{NJP.20.095002}.

In this paper, we discuss the application of optimal control theory for the fast transport of Bose-Einstein condensates with atom chips while simultaneously controlling the quantum degrees of freedom of the problem to target the ground state of the final trap as the optimization result. The article is organized as follows: We describe in the next section the chip model used to transport the BEC. The chosen cost functional and the associated transport ramp are presented in the next section, which is followed by a  comparison of our findings to the results of the STA technique applied in a similar context \cite{Corgier2018}. We finally illustrate the impact of the OCT ramp duration on the internal degrees of freedom of the final BEC state. We conclude by discussing the limits of the methodology we have developed, and by mentioning potential experimental implementations.

\section*{Chip model}

We consider the case of a Z-shaped chip configuration used to trap and manipulate cold Rb atoms in micro-gravity (See reference [\citenum{Corgier2018}] for a detailed description of the numerical model and reference [\citenum{RudolphNJP2015}] for the description of an experimental implementation). The three spatial dimensions are denoted by the three coordinates $X$, $Y$ and $z$. $z$ is the direction perpendicular to the chip. $X$ and $Y$ are two orthogonal directions in the plane of the chip. The trap is naturally rotating in the $(XY)$ plane when the physical parameters governing the trap potential change. The diagonalization of the associated Hessian matrix allows to define two new eigen-coordinates $x$ and $y$ of the trap, rotated compared to the fixed $X$ and $Y$ coordinates \cite{Corgier2018}. The physical parameters which govern the trap potential are the chip intensity $I_w$ and the bias magnetic field $B_{bias}$. For the present study, $I_w$ is fixed at 5\,A and the control parameter for the implementation of the transport ramp is the time-dependent bias magnetic field $B_{bias}(t)$, which varies between $B_{bias}(0)=B_i=21.5$\,G at the initial time $t=0$ and $B_{bias}(t_f)=B_f=4.5$\,G at the end of the transport corresponding to $t=t_f$.

In such a configuration already described in our previous study\cite{Corgier2018}, the minimum of the trap is at the origin in $x$ and $y$, and it is located at a distance $z_0(t)$ from the chip surface. At $t=0$ we have \mbox{$z_0(0) \simeq 0.45$\,mm} and at the end of the transport $z_0(t_f) \simeq 1.65$\,mm. If we limit ourselves, in a first approximation, to the simplest case of a time-dependent harmonic trap, the center-of-mass of the condensate $z_A(t)$ in the direction normal to the surface follows Newton's equations of motion
\begin{equation}
\dot{z}_A(t) = v_A(t) \quad \mathrm{and} \quad \dot{v}_A(t) = -\omega_z^2(t)\big[z_A(t)-z_0(t)\big],
\label{Eq:Newton}
\end{equation}
where $\omega_z(t)$ denotes the frequency of the trapping potential along $z$ at time $t$. Moreover, in the Thomas-Fermi approximation \cite{BECPethick2002}, the evolution of the size of the BEC is accurately described by a scaling approach \cite{PRLCastin1996,PRLKagan1997}. The size of the BEC is defined by the three time-dependent radii $r_x(t)$, $r_y(t)$ and $r_z(t)$ of the paraboloid associated with the BEC wave function, with
\begin{equation}
r_{x}(t) = r_x(0)\;\lambda_x(t),
\quad
r_{y}(t) = r_y(0)\;\lambda_y(t)
\quad \mathrm{and} \quad
r_{z}(t) = r_z(0)\;\lambda_z(t)\,.
\label{Eq:radii}
\end{equation}
It was shown \cite{PRLCastin1996,PRLKagan1997} that the time-dependent scaling factors $\lambda_x(t)$, $\lambda_y(t)$ and $\lambda_z(t)$ obey the three coupled second order differential equations
\begin{equation}
\ddot{\lambda}_x = \frac{\omega_{x}^2(0)}{\lambda_x^2\,\lambda_y\,\lambda_z}-\omega_x^2(t)\,\lambda_x,
\quad
\ddot{\lambda}_y = \frac{\omega_{y}^2(0)}{\lambda_x\,\lambda_y^2\,\lambda_z}-\omega_y^2(t)\,\lambda_y
\quad \mathrm{and} \quad
\ddot{\lambda}_z = \frac{\omega_{z}^2(0)}{\lambda_x\,\lambda_y\,\lambda_z^2}-\omega_z^2(t)\,\lambda_z\,,
\label{Eq:Castin-Dum}
\end{equation}
where $\omega_x(t)$ and $\omega_y(t)$ denote the frequencies of the trapping potential along $x$ and $y$ at time $t$. The full behavior of the trapping frequencies as a function of the control parameter $B_{bias}$ can be found in [\citenum{Corgier2018}]. Initially the trapping frequencies are $\omega_x(0) \simeq 2\pi \cdot 15$\,Hz and $\omega_y(0) \simeq \omega_z(0) \simeq 2\pi \cdot 616$\,Hz. At the end of the transport $\omega_x(t_f) \simeq 2\pi \cdot 10$\,Hz and $\omega_y(t_f) \simeq \omega_z(t_f) \simeq 2\pi \cdot 32$\,Hz. The largest time scale associated with the trap is therefore of the order of 100\,ms. An adiabatic transport would thus require transport durations larger than 1\,s. Here we want to design a simple, fast and efficient transport ramp for the BEC. The OCT technique being very powerful, we have decided to optimize a single control parameter, $B_{bias}(t)$, in order to control the final position of the BEC $z_A(t_f)$, its final speed $v_A(t_f)$, and its final size defined by the three final scaling factors $\lambda_x(t_f)$, $\lambda_y(t_f)$ and $\lambda_z(t_f)$. We also wish to control the final expansion rates given by $\dot{\lambda}_x(t_f)$, $\dot{\lambda}_y(t_f)$ and $\dot{\lambda}_z(t_f)$. Finally, since we want the harmonic approximation to hold during the entire transport, we also wish to limit the time-dependent offset between the position of the center of mass of the BEC and the center of the trap $|z_A(t)-z_0(t)|$ as well as the the time-dependent offset between their respective speeds $|v_A(t)-\dot{z}_0(t)|$. To be compatible with metrology applications with an integration over thousands of experimental cycles, we want this transport to be realized quickly, \emph{i.e.} in a duration of the order of the largest time scale associated with the trap, that is of the order of 100\,ms with the present chip configuration.

\section*{Cost functional}

To implement such an optimal control scheme, we first introduce the ``classical'' point-wise translational energy of the condensate in the reference frame of the trap
\begin{equation}
E_{cl}(t) = \frac{m}{2}\left(\omega_z^2\big[z_{A}-z_0\big]^2 + \big[v_{A}-\dot{z}_0\big]^2\right),
\label{Eq:Ecl}
\end{equation}
as well as the ``quantum'' energy of associated with the 3D Thomas-Fermi wave function
\begin{equation}
E_{qu}(t) = \frac{m}{14}\Big[\omega_x^2r_x^{\,2}+\omega_y^2r_y^{\,2}+\omega_z^2r_z^{\,2}\Big]
+ \frac{m}{14}\Big[\dot{r}_x^{\,2}+\dot{r}_y^{\,2}+\dot{r}_z^{\,2}\Big]
+ \frac{15 g N}{28\pi\,r_{x\,}r_{y\,}r_{z\,}}\,,
\label{Eq:Equ}
\end{equation}
where $g=4\pi\hbar^2a_s/m$ is the scattering amplitude, $a_s$ is the s-wave scattering length of Rb-87 and $N=10^5$ denotes the number of condensed atoms. The first term in Eq.\,(\ref{Eq:Equ}) describes the potential energy associated with the finite size of the condensate, the second term is the kinetic energy associated with the size dynamics, and the third and last term is the average mean-field interaction energy between the atoms of the condensate. The numerical factors $(1/14)$ and $(15/28)$ seen in Eq.\,(\ref{Eq:Equ}) come from the specific definition given in Eq.\,(\ref{Eq:radii}) of the size of the condensate using a Thomas-Fermi expression for the probability density.

The goal we want to achieve is the minimization of a total cost functional $C_{tot}$, defined by the sum
\begin{equation}
C_{tot} = C_{term} + C_{run}
\label{Eq:Cost}
\end{equation}
of a terminal cost
\begin{equation}
C_{term} = \lambda_1\,E_{cl}(t_f) + \lambda_2\,E_{qu}(t_f)
\end{equation}
and a running cost
\begin{equation}
C_{run} = \lambda_3\,\left(\frac{1}{t_f}\int_0^{t_f}E_{cl}(t)\,dt\right)\,.
\end{equation}
The terminal cost was designed to insure the formation of the ground state of the trap at time $t_f$. It imposes the minimization of the total energy of the condensate at the end of the transport. The running cost is introduced in order to limit the transient excitation of the condensate in the moving harmonic trap. Here we fix $\lambda_1=1$ and the two other dimensionless parameters $\lambda_2$ and $\lambda_3$ are chosen to express the relative weights between the three terms of the cost functional. Changing the values of $\lambda_2$ and $\lambda_3$ affects the progress of the optimization procedure by changing the path it takes during optimization. This can lead in practice to different final transport ramps, which will take into account the relative weight assigned to each of the terms of the cost functional.

\section*{Transport ramp}

The initial and final traps are defined by the initial and final values $B_i=21.5$\,G and $B_f=4.5$\,G of the bias magnetic field $B_{bias}(t)$. Since in experiments one can be limited by the switch on/off speed of the magnetic field  we circumvent this problem by insuring a smooth variation of $B_{bias}(t)$ at $t=0$ and at $t=t_f$. For this reason we have chosen to impose
\begin{equation}
B(t) = B_i + \big(B_f - B_i\big) \left(   10\,\left[\frac{u(t)-u_0}{u_f-u_0}\right]^3
                                        - 15\,\left[\frac{u(t)-u_0}{u_f-u_0}\right]^4
                                        +  6\,\left[\frac{u(t)-u_0}{u_f-u_0}\right]^5 \right)\,,
\label{Eq:Bbias}
\end{equation}
where $u(t)$ is a continuous function of time, with $u_0=u(0)$ and $u_f=u(t_f)$. This definition allows to impose the following boundary conditions for the bias magnetic field
\begin{equation}
\begin{array}{cclcccccl}
B_{bias}(0)        & = & B_i, & \quad &              & \quad & B_{bias}(t_f)        & = & B_f,\\[0.2cm]
\dot{B}_{bias}(0)  & = & 0,   & \quad & \mathrm{and} & \quad & \dot{B}_{bias}(t_f)  & = & 0,\\[0.2cm]
\ddot{B}_{bias}(0) & = & 0,   & \quad &              & \quad & \ddot{B}_{bias}(t_f) & = & 0\,.
\end{array}
\end{equation}
Note that a consequence of these boundary conditions imposed on $B_{bias}(t)$ is that similar relations hold for all trap parameters such has the trap position $z_0(t)$ and the trap frequencies in all directions $\omega_x(t)$, $\omega_y(t)$ and $\omega_z(t)$. The optimization procedure we have adopted is therefore using the dimensionless control function $u(t)$, from which we can calculate the optimal bias magnetic field using Eq.\,(\ref{Eq:Bbias}).

\section*{Optimal control}

We now reformulate our optimization problem in the framework of optimal control theory. We refer the interested reader to standard textbooks for details \cite{Kirk2004, Bryson1975, Bonnard2012, Alessandro2008}. The state of the system is described by a state vector $\mathbf{x}$, with
\begin{equation}
\begin{array}{cclccclccclcccl}
x_1 & = & z_A(t), & \quad & x_3 & = & \lambda_x(t),
                  & \quad & x_5 & = & \lambda_y(t),
                  & \quad & x_7 & = & \lambda_z(t)\\[0.2cm]
x_2 & = & v_A(t), & \quad & x_4 & = & \dot{\lambda}_x(t),
                  & \quad & x_6 & = & \dot{\lambda}_y(t),
                  & \quad & x_8 & = & \dot{\lambda}_z(t)
\end{array}
\end{equation}
As suggested by Eqs.\,(\ref{Eq:Newton}) and\,(\ref{Eq:Castin-Dum}), the time evolution of all components of the state vector $\mathbf{x}$ is governed by a set of coupled first order differential equations controlled by $u(t)$ through the time dependence of the trap position and frequencies. Once $u(t)$ is chosen and for well defined initial conditions at $t=0$, these equations are easily solved using a Runge-Kutta algorithm \cite{MARunge1895, ZMPKutta1901} or the Verlet method \cite{PRVerlet1967}, for instance.

According to the Pontryagin maximum principle \cite{Pontryagin1964, Lee1967}, the extremal solutions of the problem, candidates to be optimal, satisfy the equations of Hamiltonian
\begin{equation}
\dot{x}_i = +\left(\frac{\partial H_p}{\partial p_i}\right)
\quad\mathrm{and}\quad
\dot{p}_i = -\left(\frac{\partial H_p}{\partial x_i}\right),
\quad\mathrm{for\;\;} i = 1, 2, \dots, 8
\label{Eq:pi-point}
\end{equation}
where $\mathbf{p}$ is the adjoint state vector and where the Pontryagin Hamiltonian of the system is defined by
\begin{equation}
H_p(\mathbf{x},\mathbf{p},t,u) =   \mathbf{p}\bigcdot\mathbf{\dot{x}}
                                 - \lambda_3\left(\frac{E_{cl}(t)}{t_f}\right)\,.
\end{equation}

From Eq.\,(\ref{Eq:pi-point}), it can be easily shown that the dynamics of the adjoint state is governed by the following set of coupled first order differential equations
\begin{subequations}
\begin{eqnarray}
\dot{p}_1 & = & \omega_z^2(t)\left[p_2 + \lambda_3\frac{m}{t_f}\big(x_1-z_0\big)\right]\\
\dot{p}_2 & = & - p_1 + \lambda_3\frac{m}{t_f}\big(x_2-\dot{z}_0\big)\\
\dot{p}_3 & = &   p_4\left[\omega_x^2(t)+\frac{2\,\omega_{x}^2(0)}{x_{3\,}^3x_{5\,}x_7}\right]
                + p_6\left[\frac{\omega_{y}^2(0)}{x_{3\,}^2x_{5\,}^2x_7}\right]
                + p_8\left[\frac{\omega_{z}^2(0)}{x_{3\,}^2x_{5\,}x_7^2}\right]\\
\dot{p}_4 & = & - p_3\\
\dot{p}_5 & = &   p_4\left[\frac{\omega_{x}^2(0)}{x_{3\,}^2x_{5\,}^2x_7}\right]
                + p_6\left[\omega_y^2(t)+\frac{2\,\omega_{y}^2(0)}{x_{3\,}x_{5\,}^3x_7}\right]
                + p_8\left[\frac{\omega_{z}^2(0)}{x_{3\,}x_{5\,}^2x_7^2}\right]\\
\dot{p}_6 & = & - p_5\\
\dot{p}_7 & = &   p_4\left[\frac{\omega_{x}^2(0)}{x_{3\,}^2x_{5\,}x_7^2}\right]
                + p_6\left[\frac{\omega_{y}^2(0)}{x_{3\,}x_{5\,}^2x_7^2}\right]
                + p_8\left[\omega_z^2(t)+\frac{2\,\omega_{z}^2(0)}{x_{3\,}x_{5\,}x_7^3}\right]\\
\dot{p}_8 & = & - p_7\,.
\end{eqnarray}
\label{Eq:Dyn-p}
\end{subequations}
In addition, the transversality conditions for the adjoint state read
\begin{equation}
p_n(t_f) = - \lambda_1\,\left(\frac{\partial E_{cl}}{\partial x_n}\right)_{t=t_f}
           - \lambda_2\,\left(\frac{\partial E_{qu}}{\partial x_n}\right)_{t=t_f}
\end{equation}
thus leading to the following boundary conditions at time $t=t_f$
\begin{subequations}
\begin{eqnarray}
p_1(t_f) & = & - \lambda_1\,m\omega_z^2(t_f)\,\big[z_A(t_f)-z_0(t_f)\big]\\
p_2(t_f) & = & - \lambda_1\,m v_A(t_f)\\
p_3(t_f) & = & - \lambda_2\left[  \frac{m}{7}\omega_x^2(t_f)r_x(0)r_x(t_f)
                                - \frac{15gNr_x(0)}{28\pi\,r_x^2(t_f)r_y(t_f)r_z(t_f)}\right]\\
p_4(t_f) & = & - \lambda_2 \frac{m}{7}r_x(0)\dot{r}_x(t_f)\\
p_5(t_f) & = & - \lambda_2\left[  \frac{m}{7}\omega_y^2(t_f)r_y(0)r_y(t_f)
                                - \frac{15gNr_y(0)}{28\pi\,r_x(t_f)r_y^2(t_f)r_z(t_f)}\right]\\
p_6(t_f) & = & - \lambda_2 \frac{m}{7}r_y(0)\dot{r}_y(t_f)\\
p_7(t_f) & = & - \lambda_2\left[  \frac{m}{7}\omega_z^2(t_f)r_z(0)r_z(t_f)
                                - \frac{15gNr_z(0)}{28\pi\,r_x(t_f)r_y(t_f)r_z^2(t_f)}\right]\\
p_8(t_f) & = & - \lambda_2 \frac{m}{7}r_z(0)\dot{r}_z(t_f)\,.
\end{eqnarray}
\label{Eq:Bound-p}
\end{subequations}

We use a standard first-order gradient algorithm which is adapted to the control problem under study. The optimization procedure proceeds as follows:
\begin{enumerate}[label=\roman*)]
	\item First we fix an initial control ramp $u(t)$ arbitrarily, such as the linear ramp $u(t)=t/t_f$ for instance, or the STA ramp obtained from Ref. [\citenum{Corgier2018}] ;
	\item We then compute the magnetic field $B_{bias}(t)$ using Eq.\,(\ref{Eq:Bbias}) and we deduce the trap dynamics by calculating the trap motion $z_0(t)$ and the trap frequencies $\omega_x(t)$, $\omega_y(t)$ and $\omega_z(t)$ ;
	\item Using the Verlet method \cite{PRVerlet1967}, we then solve Eqs.\,(\ref{Eq:Newton}) and \,(\ref{Eq:Castin-Dum}) to simulate the condensate dynamics in the Thomas-Fermi regime from the initial time $t=0$ to the final time $t=t_f$ ;
	\item We calculate the adjoint state $\mathbf{p}(t_f)$ at the end of the transport using Eq.\,(\ref{Eq:Bound-p}) and we propagate $\mathbf{p}(t)$ backward in time until $t=0$ using Eq.\,(\ref{Eq:Dyn-p}) ;
	\item Finally, we add a first order correction to the control ramp by replacing the control function $u(t)$ by $[u(t)+\delta u(t)]$, where $\delta u(t) = \epsilon(\partial H_p/\partial u)$, $\epsilon$ being a small positive constant.
\end{enumerate}
This procedure is repeated until convergence is reached.

\section*{Convergence}

\begin{figure}[t]
	\centering
	\includegraphics[width=10cm]{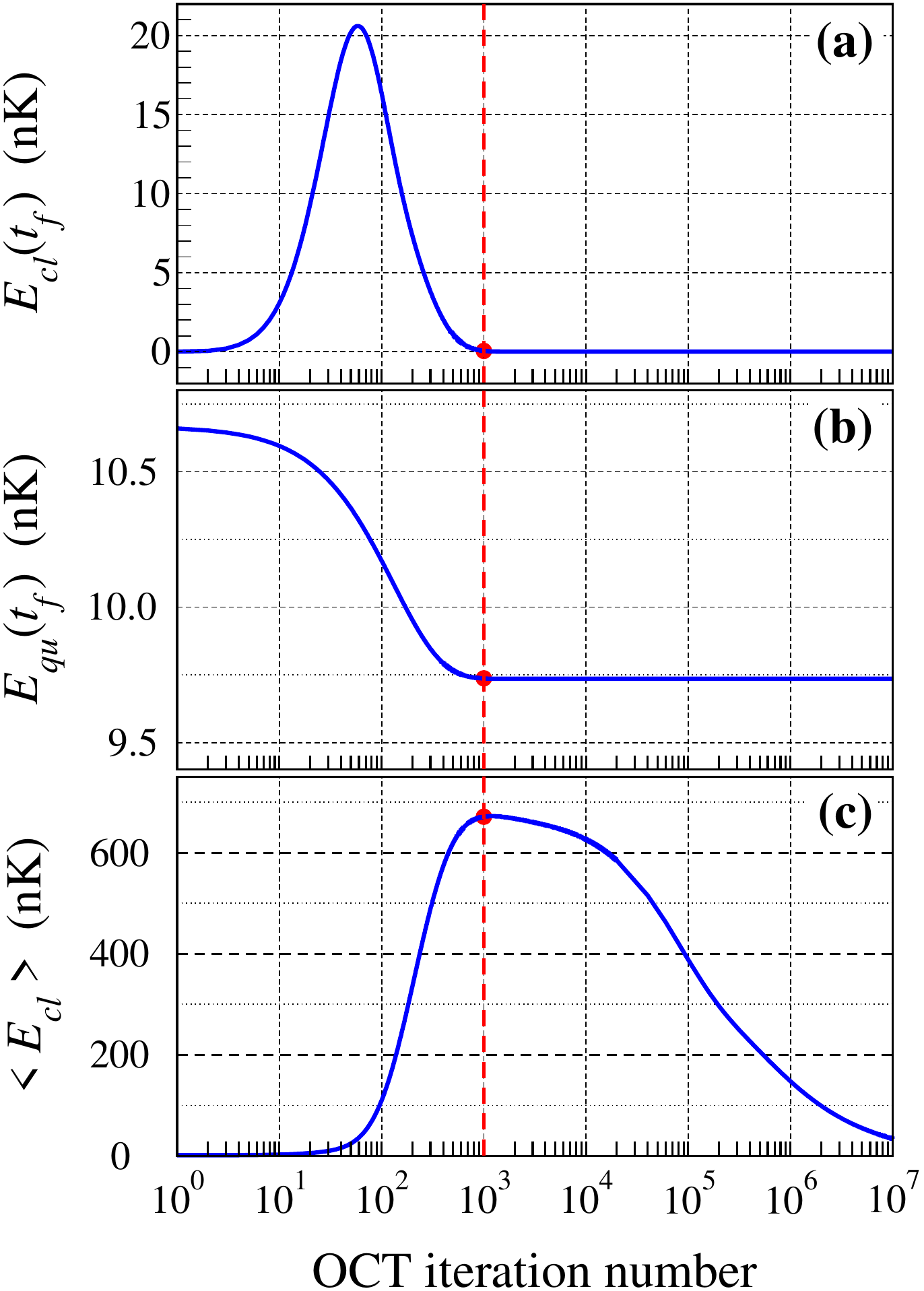}
	\caption{Example of convergence of the different cost functionals as a function of the optimal control theory iteration number: (a) Final classical energy in nK, (b) Final quantum energy in nK, (c) Average classical energy in nK. See text for details.}
	\label{fig:convergence}
\end{figure}

Figure \ref{fig:convergence} shows a typical example of convergence of this algorithm. The condensate is assumed to be initially at rest in the ground state of the initial trap. The initial control ramp is the shortcut-to-adiabaticity solution (see Ref.\,[\citenum{Corgier2018}] for details). The weight parameters are $\lambda_1=1$, $\lambda_2=5.10^{5}$ and $\lambda_3=0.001$. We have chosen in this example a large value for $\lambda_2$ in order to impose a fast convergence for the control of the final size of the condensate. In practice the correction parameter $\epsilon$ has to be chosen small enough to insure the convergence of the optimization algorithm. Since the correction to the control ramp is introduced at first order only, decreasing the value of $\epsilon$ beyond a reasonable limit does not improve the accuracy of the optimization procedure but it slows down the convergence. In the present example we have chosen $\epsilon=10^{-11}$. In Fig. \ref{fig:convergence}, panel (a) shows the classical energy $E_{cl}(t_f)$ of the condensate at the end of the transport ($t_f=150$\,ms in this case) as a function of the optimal control theory iteration number (logarithmic scaling). Panel (b) shows the quantum energy $E_{qu}(t_f)$ of the condensate at the end of the transport as a function of the iteration number. Panel (c) shows the average classical energy $\frac{1}{t_f}\int_0^{t_f}E_{cl}(t)dt$ of the condensate during the transport as a function of the iteration number.

Since the total cost functional given in Eq.\,(\ref{Eq:Cost}) is characterized by a very large weight $\lambda_2$ associated with the final quantum energy, we see that $E_{qu}(t_f)$ is very quickly minimized, in about 1,000 iterations. This limit of 1,000 iterations is emphasized in Figure \ref{fig:convergence} with a vertical dashed red line. Once this convergence is reached, the final 3D size of the condensate adopts the size of the ground state of the final trap and the size dynamics of the BEC is frozen. This convergence was obtained at the cost of a transient degradation of the final classical energy, which reaches a maximum of about 20\,nK after about 60 iterations, but the final classical energy is then minimized very quickly to reach a near-zero value in about 1000 iterations. It is only when this first stage of convergence is reached (iteration number $> 1000$) that the last cost functional, associated with a smaller weight $\lambda_3$, starts to decrease. One can note that the convergence of the average classical energy during the transport [in panel (c)] is rather slow since it requires more than $10^7$ iterations before it starts to stabilize at values close to 30\,nK. This value can be compared with the energy of the condensate in the initial trap, which is close to 120\,nK, and with the energy of the condensate in the final trap, close to 10\,nK. The transient excitation during the transport is therefore relatively limited.

\begin{figure}[t]
	\centering
	\includegraphics[width=13cm]{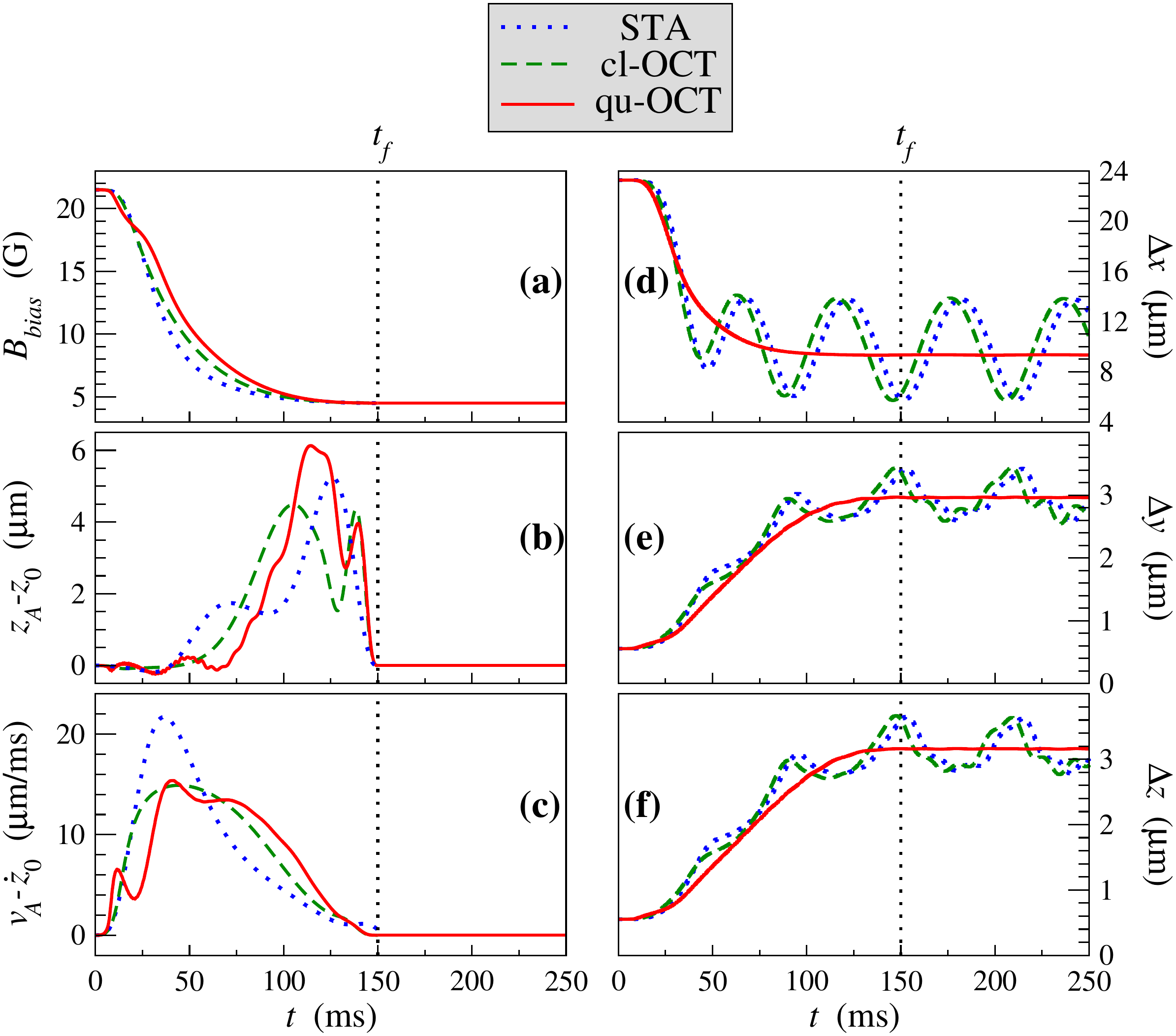}
	\caption{Comparison of different optimization procedures. Shortcut-to-adiabaticity (STA): dotted blue line, classical optimal control (cl-OCT): dashed green line, quantum optimal control (qu-OCT): solid red line. (a) Bias magnetic field in Gauss as a function of time, (b) Position offset $[z_A(t)-z_0(t)]$ in $\mu$m as a function of time, (c) Velocity offset $[v_A(t)-\dot{z}_0(t)]$ in $\mu$m/ms as a function of time, (d)-(f) Size dynamics of the condensate along the three coordinates $x$, $y$ and $z$ in $\mu$m as a function of time. The duration of the transport is $t_f=150$\,ms. See text for details.}
	\label{fig:comparison}
\end{figure}

\section*{Comparison of different optimization procedures}

In figure \ref{fig:comparison} the Shortcut-To-Adiabaticity (STA) transport ramp obtained in Ref.\,[\citenum{Corgier2018}] (dotted blue line) is compared with two results obtained with the present optimal control technique (OCT). Note that strictly speaking, these two methods assume slightly different constraints and that, in principle, STA can be combined with the minimization of a cost functional. However, a quantitative study on this specific point is beyond the scope of this article where we concentrate mainly on developing the transport method with the OCT protocol. The correction parameter is $\varepsilon=10^{-10}$. The dashed green line labeled as ``cl-OCT'' shows the result obtained for the weight factors $\lambda_1=1$, $\lambda_2=0$ and $\lambda_3=5.5\,10^{-4}$. The solid red line labeled as ``qu-OCT'' is for $\lambda_1=1$, $\lambda_2=3.3$ and $\lambda_3=5.5\,10^{-4}$. The difference between these two OCT results lies in the fact that qu-OCT takes into account the influence of the finite size of the BEC in the cost functional, while cl-OCT considers the BEC as a classical point-wise particle. The BEC model used for cl-OCT is therefore similar to the model used in STA and these two approaches can be compared directly. The optimized time variation of the bias magnetic field $B_{bias}(t)$ is shown as a function of time in the first panel (a). The duration of the transport is $t_f=150$\,ms, and all results are plotted from $t=0$ to $t=250$\,ms \emph{i.e.} up to 100\,ms after the end of the transport. This time interval was chosen in order to detect the eventual presence of a residual excitation at the end of the transport. The position $[z_A(t)-z_0(t)]$ and velocity $[v_A(t)-\dot{z}_0(t)]$ offsets are shown in panels (b) and (c). Finally, Panels (d), (e) and (f) present the condensate size dynamics $\Delta\alpha(t)$ along the three coordinates $\alpha \equiv x, y$ or $z$, where $\Delta\alpha(t)=r_\alpha(t)/\sqrt{7}$ represents the width (standard deviation) of the Thomas-Fermi condensate wave function in the directions $\alpha \equiv x, y$ or $z$.

We see in panels (b) and (c) that the three methods are very efficient for the control of the final average position and velocity of the BEC since the condensate is fully at rest in the center of the trap at the end of the transport and for all times $t > t_f = 150$\,ms. In addition, the transient position and velocity offsets during the transport reach similar values using these three different optimization methods. One can note in panels (b) and (c) that in terms of maximum transient offset in position and speed, from the two methods that we can compare directly, cl-OCT is a little better than STA (maximum offsets of 4.5\,$\mu$m \emph{vs.} 5.3\,$\mu$m in position and 14\,$\mu$m/ms \emph{vs.} 22\,$\mu$m/ms in speed) but this difference is not very significant in practice. The transient offsets of the qu-OCT approach are slightly larger than those of the cl-OCT method (with maximum offsets of 6.2\,$\mu$m in position and 15\,$\mu$m/ms in speed). Again this increase would be very benign in a practical implementation. Note finally that the three control fields $B_{bias}(t)$ shown in panel (a) are relatively similar, with a fast initial decrease during the first half of the ramp, before 75\,ms, followed by a much slower decrease afterward. A first conclusion of this study is therefore that, if one is mainly interested in the control of the average translational degree of freedom of the BEC, the STA approach, whose numerical implementation is much simpler than OCT, is sufficient.

\begin{figure}[t]
	\centering
	\includegraphics[width=17cm]{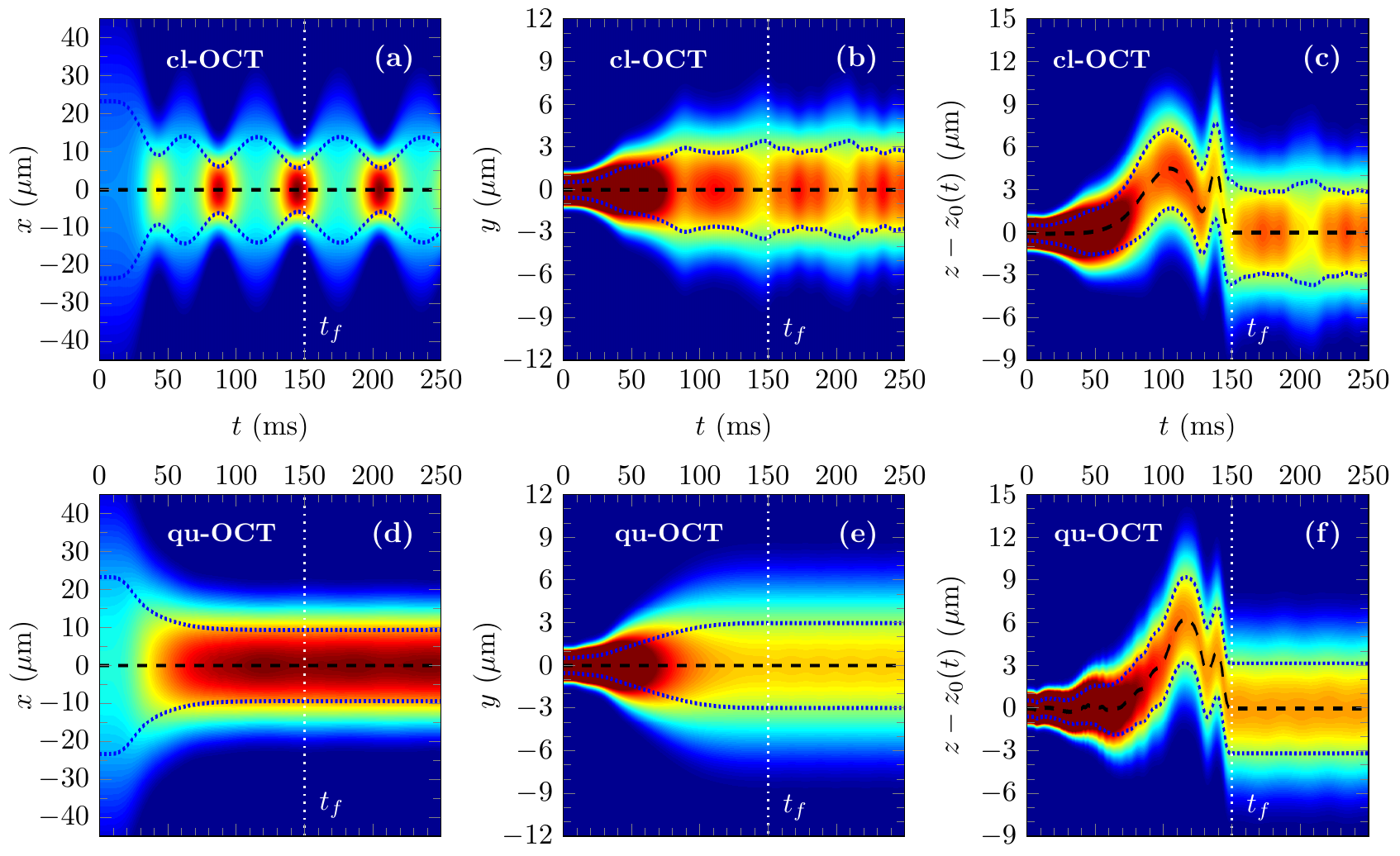}
	\caption{Condensate dynamics in the $x$, $y$ and $z$ directions using the cl-OCT ramp (upper line) and the qu-OCT ramp (lower line) shown in Figure \ref{fig:comparison}. The transport duration is $t_f=150$\,ms. The average atomic density, solution of the time-dependent Gross-Pitaevskii equation, is shown as a function of time and position: (a) and (d) $P_x(x,t)$, (b) and (e) $P_y(y,t)$, (c) and (f) $P_z(z,t)$. The black dashed lines show the expected center of mass trajectory. The dotted blue lines highlight the expected width dynamics according to the scaling approach. The dotted vertical white lines mark the time of the end of the transport. The total atom number is $N=10^5$. See text for details.}
	\label{fig:size-dynamics}
\end{figure}

It is in the size dynamics shown in panels (d), (e) and (f) that there is a striking difference between qu-OCT and the two other optimization methods. In terms of size dynamics, cl-OCT and STA give very similar results which consist in a persistent size excitation of the condensate after the transport. This result was already seen in Ref.\,[\citenum{Corgier2018}], where it was shown that it was mainly the first quadrupole mode $Q_1$ which was excited, thus explaining that the size oscillation along $x$, $y$ and $z$ is almost periodic after the transport. The qu-OCT approach is able to suppress efficiently this quadrupole-mode excitation and, at the end of the transport, the sizes $\Delta x$, $\Delta y$ and $\Delta z$ remain constant. We can therefore conclude that the introduction of a minimization goal for the quantum energy associated with the finite size dynamics of the condensate allows the qu-OCT transport ramp to prepare the true ground state of the final trap at $t=t_f$. When the size dynamics is not accounted for, as in the STA and cl-OCT approaches, it is impossible to insure the preparation of the lowest energy state in the final trap using short transport ramps.

The optimized OCT transport ramps were obtained using a Thomas-Fermi approximation in a 3D harmonic trap. We have therefore verified, by solving the 3D mean-field time-dependent Gross-Pitaevskii equation for the evolution of the time-dependent macroscopic condensate wave function $\psi(x,y,z,t)$, that this control is robust when taking into account the anharmonicities and the rotation of the trap. The numerical method used for this calculation is described in Ref. [\citenum{Corgier2018}]. This result is illustrated in Figure \ref{fig:size-dynamics}, showing the time evolution of the average atomic densities
\begin{subequations}
	\begin{eqnarray}
	P_x(x,t) & = & \int_{-\infty}^{\infty}dy\int_{-\infty}^{\infty}dz\;|\psi(x,y,z,t|^2\\
	P_y(y,t) & = & \int_{-\infty}^{\infty}dx\int_{-\infty}^{\infty}dz\;|\psi(x,y,z,t|^2\\
	P_z(z,t) & = & \int_{-\infty}^{\infty}dx\int_{-\infty}^{\infty}dy\;|\psi(x,y,z,t|^2
	\end{eqnarray}
\end{subequations}
along $x$, $y$ and $z$. In each panel the black dashed line shows the expected center of mass trajectory obtained by solving Newton's equations of motion\,(\ref{Eq:Newton}). Similarly, the dotted blue lines highlight the expected widths obtained by solving the scaling equations\,(\ref{Eq:Castin-Dum}). The condensate wave function follows clearly these predicted positions and widths. It is therefore clear from Figure \ref{fig:size-dynamics} that the controls predicted by the cl-OCT and qu-OCT methods are robust with respect to the anharmonicities and with respect to the inherent rotation of the trap in this realistic atom chip setup. In addition, the control of collective excitations using the qu-OCT approach appears clearly when comparing the lower line (qu-OCT ramp) of Figure \ref{fig:size-dynamics} with the upper line (cl-OCT ramp) of the same Figure.

\begin{figure}[ht]
	\centering
	\includegraphics[width=12cm]{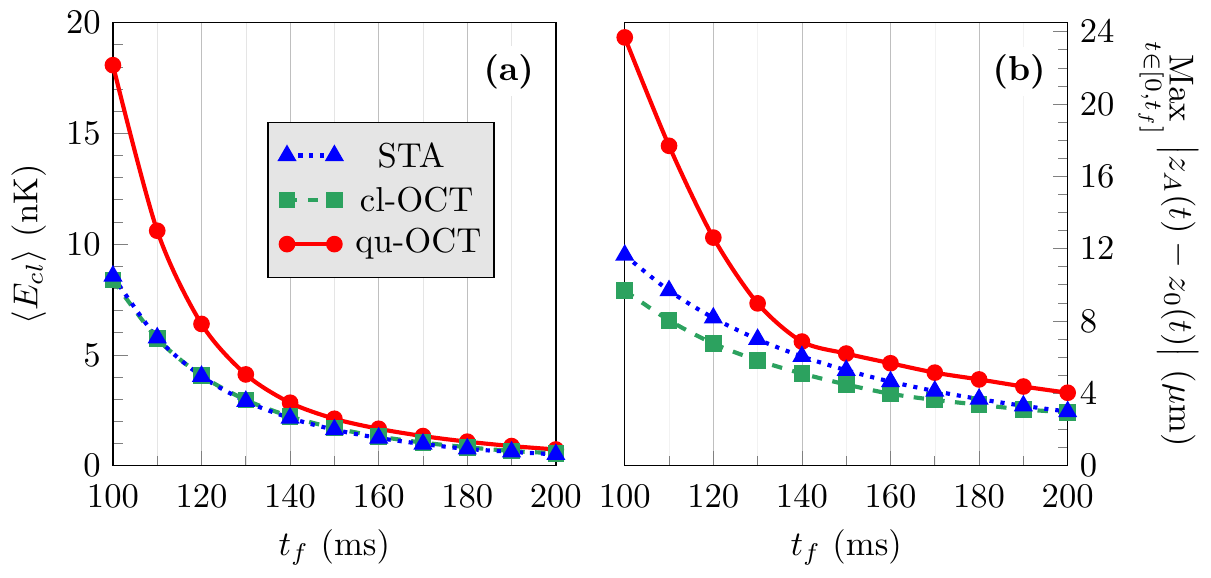}
	\caption{Influence of the transport duration $t_f$ on: (a) the average translational energy $\langle E_{cl} \rangle$ of the condensate and (b) the maximum position offset $|z_A-z_0|$ during the transport. Shortcut-to-adiabaticity (STA): dotted blue line, classical optimal control (cl-OCT): dashed green line, quantum optimal control (qu-OCT): solid red line. The weight parameters $\lambda_1$, $\lambda_2$ and $\lambda_3$ are the same as those used in Fig. \ref{fig:comparison}. See text for details.}
	\label{fig:transp-duration1}
\end{figure}

\section*{Influence of the transport duration}

What remains to be seen is the efficiency of these various optimization procedures for different transport durations. Figure \ref{fig:transp-duration1} shows in panel (a), for the three optimized ramps, the variation of the average translational energy
\begin{equation}
\langle E_{cl} \rangle = \frac{1}{t_f}\int_0^{t_f}E_{cl}(t)\,dt
\end{equation}
as a function of the ramp duration $t_f$. Panel (b) shows, in the same conditions, the maximum position offset $\mathrm{Max}\,\big|z_A(t)-z_0(t)\big|$ during the transport. We see here that whatever the transport duration STA and cl-OCT are characterized by a very similar performance in terms of transient excitations. This confirms the advantage of the STA protocol in practical implementations, due to its overall simplicity when compared to cl-OCT. We also see that, on one hand, when the transport duration is larger than 140\,ms (\emph{i.e.} about 1.4 times the largest time scale associated with the trap), the transient excitations realized by the improved qu-OCT procedure are very close to the ones of cl-OCT and STA. On the other hand, for transport durations smaller than 140\,ms larger transient excitations are obtained when using qu-OCT.

\begin{figure}[ht]
	\centering
	\includegraphics[width=17cm]{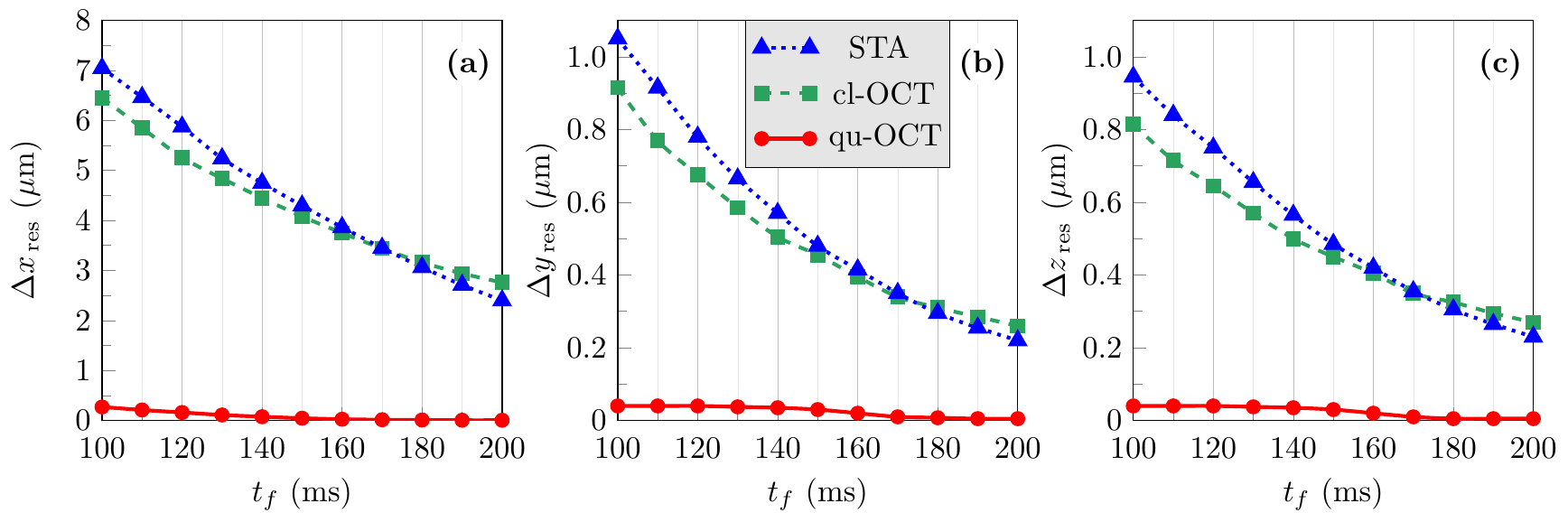}
	\caption{Residual oscillation amplitudes in the size dynamics after transport in the (a) $x$, (b) $y$ and (c) $z$ directions, as a function of the transport duration $t_f$. Shortcut-to-adiabaticity (STA): dotted blue line, classical optimal control (cl-OCT): dashed green line, quantum optimal control (qu-OCT): solid red line. The weight parameters $\lambda_1$, $\lambda_2$ and $\lambda_3$ are the same as those used in Fig. \ref{fig:comparison}. See text for details.}
	\label{fig:transp-duration2}
\end{figure}

We could however verify that for all transport durations in the range $100\,\mathrm{ms} \leqslant t_f \leqslant 200\,\mathrm{ms}$, the qu-OCT method is able to minimize very efficiently the residual size excitations after the transport, a goal which is not achievable with the STA or cl-OCT procedures. This can be seen in Figure \ref{fig:transp-duration2}, which shows the residual oscillation amplitudes
\begin{equation}
\Delta\alpha_{\,\mathrm{res}} = \frac{1}{2}\,\Big[\, \underset{t\,>\,t_f}{\mathrm{Max}}\big(\Delta\alpha\big)
                               - \underset{t\,>\,t_f}{\mathrm{Min}}\big(\Delta\alpha\big)\,\Big]
\end{equation}
of the size of the condensate $\Delta\alpha$ (standard deviation of the Thomas-Fermi condensate wave function) after the transport, for $\alpha \equiv x$ [panel (a)], $\alpha \equiv y$ [panel (b)], and $\alpha \equiv z$ [panel (c)]. The results shown in Figures \ref{fig:transp-duration1} and \ref{fig:transp-duration2} demonstrate that for $t_f \geqslant 140$\,ms the residual size excitations of the condensate can be limited efficiently by optimal control and that this limitation does not introduce any detrimental effect on the transient excitation of the BEC. The same result can also be obtained by optimal control with shorter transport ramps, but at the cost of an increased transient excitation of the condensate.

\section*{Conclusion}

In conclusion, we engineered optimal control theory protocols allowing for the fast, excitation-less transport of BECs over large distances compatible with a precision atom interferometric use. The ramps presented in this work relied on a single-parameter (bias magnetic field) optimization to shift the trap minimum position of the atom chip, promising a straightforward experimental implementation. The results of the OCT procedure relied on a scaling approach assuming a harmonic trapping. Real-life implementations on atom chips comes with anharmonic corrections, mainly cubic in the direction of the transport, that scale with the position offset between the atoms and the trap minimum during the transport and with an inherent rotation of the trap. We demonstrated in this study, by solving 3D Gross-Pitaevskii equations for typical anharmonic and rotating chip traps, that the proposed OCT protocol does not compromise the target state solution even for very competitive ramp times of 150\,ms. This also suggests a successful transfer to experiments. Moreover, we indicated by studying the impact of different transport durations, the methodology to follow in order to device the shortest ramps possible. Indeed, by quantifying the maximum offset induced by each ramp duration, every experimental implementation would be characterized by an anharmonicity range explored according to the specific trap configuration considered. This range determines, ultimately, the success of the ramp in reaching the ground state of the final trap. The positive outcome of this study suggests a natural generalization to the dual-species transport case. For this latter, no analytic neither intuitive solutions do exist. The STA approach generally fails since the two species experience different potential frequencies due their mass difference. A comparable OCT approach to the one adopted in this study, based on a pair of coupled mean-field equations, would allow to find trap trajectories that bring a quantum mixture to a target position in its ground state. Such a source will allow for precision interferometric measurements such as equivalence principle tests.

\section*{Acknowledgements}

This work is supported by the DLR Space Administration with funds provided by the Federal Ministry of Economics and Technology (BMWi) under grant numbers DLR 50WM1131-1137, 50WM0940 and 50WM1240. N.G. acknowledges financial support of the “Nieders\"achsisches Vorab” through the “Quantum- and Nano-Metrology” (QUANOMET) initiative within the project QT3. S.A. and R.C were supported by mobility scholarships of the DAAD. S.A., N.G. and E.C. would like to acknowledge networking support by the COST action CA16221 “Atom Quantum Technologies”. R.C. and N.G. acknowledge mobility support from the DAAD Procope action, from the IP@Leibniz program of the LU Hanover and from the Q-SENSE project funded by the European Union’s Horizon 2020 Research and Innovation Staff Exchange (RISE) under Grant Agreement Number 691156. N.G. and E.M.R. acknowledge support from the DFG SFB 1227 DQmat. We also acknowledge the use of the computing cluster MesoLum/GMPCS of the LUMAT research federation (FR 2764 CNRS). The authors thank Waldemar Herr, David Guéry-Odelin and Holger Ahlers for valuable discussions.

\section*{Author contributions statement}

SA derived the OCT model, developed and performed the OCT and GPE numerical simulations, and prepared the figures. RC produced and delivered the STA ramp and helped with the interpretation of the results. DS proposed the OCT model and helped for its implementation. EMR, NG and EC designed the research goals. DS, NG and EC wrote the paper. All authors reviewed the results and the paper, and approved the final version of the manuscript.

\section*{Additional information}

The authors declare no competing interests.

\end{document}